\documentclass[reqno, centertags]{amsart}
\usepackage{amssymb}
\usepackage{graphicx}

\theoremstyle{definition}

\begin{document}

\title{Physical Realities from Simple \\Philosophical Conjectures.}

\author{Helen Lynn}

\address{Quantum Philosophy Theories, www.qpt.org.uk }

\email{helen.lynn@qpt.org.uk}

\author{Michele Caponigro}

\address{Physics Department, University of Camerino, I-62032 Camerino, Italy }

\email{michele.caponigro@unicam.it}

\date{\today}

\begin{abstract}
We will focus on the Quantum theory and starting from simple
philosophical conjectures, we infer possible different physical
realities. Also we argue of possible wavefunction emerging under
specific conditions of the physical reality. Finally, we affirm
that the "hidden choice" of the ontic elements as primitive is a
fundamental step to analyze the construction of any theory.

\end{abstract}

\maketitle
\section{Three basic elements.}

We work with three basic elements:
\begin{enumerate}
\item The observer.
\item The law of physics.
\item Physical Phenomenon.
\end{enumerate}

We assume that physical reality and physical phenomenon are
different concepts, latter is subjects of observation. Above three
elements and quantum theory will be used to infer possible
underlying physical realities.\\

\textbf{Premise:}
\begin{itemize}
\item symbol: \textbf{1} mean an \textbf{ontic} element
\item symbol: \textbf{0} mean an \textbf{epistemic} element
\item the choice of symbol "1" mean that the correspondent element is take as
primitive.
\end{itemize}

This \textbf{classical} table summarize possible realities: \\

\begin{table}[h]
\centering
\begin{tabular}{l|c|r|l}\hline

 Observer & Law Phys.& Phen. & \Large\textbf{Physical Realities}\\ \hline
 0       &  0       & 0       &  \emph{\textbf{No reality}}\\ \hline
 0       &  0       & 1       & \emph{\textbf{Realism/Emperic.}}\\ \hline
 0       &  1       & 0       & \emph{\textbf{Platonism (e.g.Rovelli\cite{Rov1}/Everett Inter.)}}\\ \hline
 0       &  1       & 1       &  \emph{\textbf{Weak Realism}}\\ \hline
 1       &  0       & 0       & \emph{\textbf{Idealism (e.g.Fuchs\cite{Fuchs1}Inter.)}}\\ \hline
 1       &  0       & 1       & \emph{\textbf{Mind/Matter Int.}}\\ \hline
 1       &  1       & 0       &  \emph{\textbf{Weak Idealism}}\\ \hline
 1       &  1       & 1       & \emph{\textbf{Monism (i.e.Anomalous Monism\cite{Davi1})}}\\ \hline
\end{tabular}
\end{table}

\textbf{Observation}: The hidden choice of the primitive element
is fundamental to determine the underlying reality. Now,the
problem is to establish if: (i)the "choice" is a
\textbf{consequence} of theory,a consequence of his experimental
data or (ii) an \textbf{"a priori"} hidden choice. We argue, that
we can find in the \textbf{premise} of any theory the
questions/answers and their consequence, in other words, we have
the start/end in the premise. In order to justify these strong and
debated statements, we can see (above table), that starting from
different ontic elements, related to the same phenomenon ( e.g.
measurement process in quantum theory) we have two different
interpretations of \textbf{underlying physical reality.}

For instance, the \textbf{first} case:
\begin{enumerate}
\item The observer is the \textbf{ontic} element
\item The law of physics are secondary
\item Physical Phenomenon is secondary
\end{enumerate}

The consequences of the measurement process are:

\begin{enumerate}
\item "Measurement" is only an information of the observer.
\item The physical reality do not exist.
\item Physical Phenomenon is provoked by observer.
\end{enumerate}

For instance, a \textbf{second} case:
\begin{enumerate}
\item Physical Phenomenon is ontic.
\item The observer is a secondary property.
\item The law of physics are secondary.
\end{enumerate}

The consequences related to measurement process are:

\begin{enumerate}
\item Physical Phenomenon is the only objective elements.
\item The observer is a reduction of the phenomenon, they have the
same properties.
\item Instrumentalist interpretation of the physical's laws.
\end{enumerate}

As we can see, the conclusions on the conception of reality are
completely different.

\section{Wavefunction inferred.}

We conclude the brief paper, with another table, now we include in
the table as variable ontic/epistemic the physical reality in
order to infer the correspondent quantum mechanics theory (i.e.
wavefunction). In this case we try to give an \textbf{"a priori"}
possible nature of physical reality.

\begin{table}[h]
\centering
\begin{tabular}{l|c|r|l}\hline
 Observer & Phys.Real.& Phen. & \textbf{\Large wavefunction\textbf{$\mid\Psi>$ of}}\\ \hline
 0       &  0       & 0      &  \emph{\textbf{$\mid\Psi>$ unknown interpretation}}\\ \hline
 0       &  0       & 1      & \emph{\textbf{$\mid\Psi>$ (e.g.GRW inter.)}}\\ \hline
 0       &  1       & 0      & \emph{\textbf{$\mid\Psi>$ (e.g.Rovelli inter.)}}\\ \hline
 0       &  1       & 1      &  \emph{\textbf{$\mid\Psi>$ (e.g.Bohm inter.)}}\\ \hline
 1       &  0       & 0      & \emph{\textbf{$\mid\Psi>$ (e.g.Copenhagen,Fuchs-Zeilinger inter.)}}\\ \hline
 1       &  0       & 1      & \emph{\textbf{$\mid\Psi>$ (e.g.Informational inter.)}}\\ \hline
 1       &  1       & 0      &  \emph{\textbf{$\mid\Psi>$ unknown interpretation}}\\ \hline
 1       &  1       & 1      & \emph{\textbf{$\mid\Psi>$ unknown interpretation}}\\ \hline
\end{tabular}
\end{table}

\section{Conclusion.}

Simple and intuitive conjectures from philosophical point of view,
maybe help us to clear the "entangled" world of the
interpretations of QM. The "a priori" choice of the ontic elements
remain a fundamental step to interpret and analyze not only
quantum theory.


\begin{thebibliography}{99}
\bibitem{Rov1}
C Rovelli: "Relational Quantum Mechanics", International Journal
of Theoretical Physics, 35, 1637 (1996)
\bibitem{Fuchs1}
C.Fuchs Quantum Mechanics as Quantum Information(and only a little
more)- http://arxiv.org/abs/quant-ph/0205039
\bibitem{Davi1}
D.Davidson: Subjective, Intersubjective, Objective Oxford:
Clarendon Press, 2001

\end{thebibliography}
\end{document}